\documentclass[pra, showpacs, a4paper, twocolumn, 10pt]{revtex4} 

\usepackage[utf8]{inputenc}
\usepackage[T1]{fontenc}
\usepackage[english]{babel}
\usepackage{bbm, amsmath, amssymb, amsthm, bm, textcomp, nicefrac}

\usepackage{graphicx}
\usepackage{geometry}
\usepackage{color}
\usepackage{float}

\addto\captionsenglish{}

\begin{document}

\title{Propagation of numerical noise in particle-in-cell tracking}

\author{Frederik Kesting}
\affiliation{IAP, Goethe University, D-60438 Frankfurt am Main, Germany}

\author{Giuliano Franchetti}
\affiliation{GSI, D-64291 Darmstadt, Germany}

\date{\today}

\begin{abstract}  
Particle-in-cell (PIC) is the most used algorithm to perform self-consistent 
tracking of intense charged particle beams. 
It is based on depositing macro-particles on a grid, and subsequently solving on it 
the Poisson equation. 
It is well known that PIC algorithms occupy intrinsic limitations as 
they introduce numerical noise. 
Although not significant for short-term tracking, 
this becomes important in simulations for circular machines over millions 
of turns as it may induce artificial diffusion of the beam. 

In this work, we present a modeling of numerical noise induced by PIC 
algorithms, and discuss its influence on particle dynamics.
The combined effect of particle tracking and noise created by PIC algorithms leads 
to correlated or decorrelated numerical noise.
For decorrelated numerical noise we derive a scaling law for the simulation parameters, allowing an estimate of artificial emittance growth. 
Lastly, the effect of correlated numerical noise is discussed, and a mitigation 
strategy is proposed.
\end{abstract}


\maketitle

\section{Introduction}\label{ch:Introduction}
It is well known that in operational scenarios requiring long-term storage, 
sources of noise in the machine can lead to detrimental effects on the beam. 
Examples include RF voltage noise, beam-beam interaction, ground motion, and feedback noise, which all have been subject of studies to mitigate or exploit their impact, as reported in Refs.~\cite{realNoise1,realNoise2,realNoise3,realNoise4,realNoise5,realNoise6}

Recently, a similar concern has raised for numerical noise in self-consistent simulation 
of high intensity beams. In fact, the study of space charge effects via self-consistent simulations has 
become important with the advent of new projects, like the future SIS100 
synchrotron of the FAIR project~\cite{FAIR}, and the LIU project~\cite{LIU} for the CERN 
accelerator complex. In these projects some scenarios require the storage of a 
high intensity bunched beam for seconds.

The simulation of these operational scenarios significantly raises the 
computational challenges, provided that they concern the prediction of beam loss, 
or emittance growth. 
In order to avoid any artificial noise, the first simulation studies 
were limited to frozen models, which rely on noise-free 
tracking schemes~\cite{frozen, frozen1}. 
There, the space charge is computed assuming the beam remains frozen, 
therefore allowing an analytic description of the space charge force. 
This approach surely avoids artificial emittance growth, but unfortunately 
at the expenses of self-consistency, which becomes relevant in certain scenarios e.g. large 
beam loss, or large beam core growth.

On the other hand, PIC simulations of high intensity beams allow a 
self-consistent treatment, but require an understanding of 
the origins and propagation of its numerical noise. Hence, simulation 
parameters can be set adequately in order to mitigate noise 
induced artifacts, while keeping the computational load feasible. 

The evaluation of PIC induced artifacts on particle tracking 
has been studied by several authors from several points of view, 
e.g. in the context of a Fokker-Planck approach, as reported
in Refs.~\cite{Struckmeier2000,Oliver2014,HofmannGrid}. 

Differently from previous approaches, we discuss the effect of PIC 
induced noise by following the integration method of a typical 
beam dynamics code. We start with an analysis of the noise due to the PIC algorithm, 
and discuss how it propagates via the beam dynamics integration. 
Following this approach, the dependence of rms-emittance growth on
simulation parameters is derived. 
It is also found that the particle's rotation in phase space creates 
correlations in numerical noise, which enhances artificial emittance growth. 

Our results are of interest for simulations of millions of turns in circular 
machines, where the high intensity requires a self-consistent modeling, hence the use of PIC algorithms, 
while at the same time the control of artificial emittance growth.

This paper is organized as follows. In Section~\ref{Sec:scaling_law} we study the 
origins and properties of static PIC noise. Then, in Sec.~\ref{ch:SingleParticle}, we 
model the effect of random PIC noise on the dynamics of a single particle, 
while the effect on the whole ensemble of particles is treated 
in Sec.~\ref{ch:Distribution}. In Section~\ref{ch:Correlations} we introduce 
the concept of correlated and decorrelated numerical noise. 
We conclude our studies with a summary in Sec.~\ref{ch:Conclusion}. 
The effect of the particle's rotation in phase space on numerical noise 
is discussed in Appendix~\ref{ch:Appendix}, whereas the consequences of the 
periodic random walk on the diffusion of macro-particles is treated in 
Appendix~\ref{ch:Appendix_RW}. Finally, in Appendix~\ref{ch:Appendix_SR_Tune}, we discuss the impact of the strength 
of space charge forces on the excitation of stochastic resonances.

\section{Scaling law for electric field fluctuations}\label{Sec:scaling_law}
Before treating the propagation of numerical noise, we study how the PIC scheme is 
causing numerical noise in a static scenario.
We start by considering a Gaussian beam, i.e. the particle's phase space 
positions are defined randomly according to a Gaussian probability 
density function (p.d.f.), given by
\begin{equation}\label{eq:GaussDistribution}
  f(x,x^\prime,y,y^\prime)=\frac{e^{-\frac{1}{2}\left(\frac{x^2}{\sigma_x^2}+\frac{x^{\prime 2}}{\sigma_x^{\prime 2}}+\frac{y^2}{\sigma_y^2}+\frac{y^{\prime 2}}{\sigma_y^{\prime 2}}\right)}}{4\pi^2\sigma_x\sigma_{x^\prime}\sigma_y\sigma_{y^\prime}},
\end{equation}
where $\sigma_x,\sigma_{x^\prime}, \sigma_y,\sigma_{y^\prime}$ are 
the transverse standard deviations. 
In the following discussion, we consider round Gaussian beams with 
$\sigma_r^2:=2\sigma_x^2=2\sigma_y^2 $ the standard deviation for 
the radial coordinate $r^2=x^2+y^2$. The associated radial space 
charge field at a longitudinal position $s_j$ is given by
\begin{equation}
  E_{r}(x,y,s_j)= \frac{E_g(s_j)}{\sqrt{x^2+y^2}}\left(1-e^{-\frac{x^2+y^2}{\sigma_r^2(s_j)}}\right),
\end{equation}
where $E_g(s_j)$ is proportional to the gradient of the electric field at $x=y=0$. 

The transverse electric field of a coasting  beam, that 
determines the self-consistent space charge forces, is computed in 
long-term simulations most efficiently using a two-dimensional 
particle-in-cell (PIC) scheme. To achieve a high computational 
efficiency, the Poisson equation in the transverse plane is solved on 
a finite set of $N_G\times N_G$ grid points and the particle distribution 
is approximated by $N_M$ macro-particles. 
Since the number of macro-particles $N_M$ is much smaller than the number 
of physical particles $N_p$, an artificial granularity of the distribution 
is introduced, which is one source of numerical noise. 

To study the impact of the simulation parameters on numerical noise, 
we adopt the random start technique (opposite to the quiet start). 
In the random start approach the positions of $N_M$ macro-particles are 
randomly initialized according to a p.d.f, as given by 
Eq.~\ref{eq:GaussDistribution} for Gaussian beams. Therefore, the resulting 
electric field is not only determined by the p.d.f. 
itself, but varies according to the random initialization of 
macro-particles with an amplitude $\delta E_x$. 
This amplitude of fluctuations $\delta E_x$ is identified with 
the standard deviation of the electric field for multiple random 
initializations of macro-particles.  

Previous studies, e.g. Refs.~\cite{PlasmaBook1,PlasmaBook2,PIC1,PIC2,PIC3}, 
have found that the parameters of relevance for a noise analysis of a beam with a 
fixed size are  $N_G$ and $N_M$, since they fix the number of particles per cell. 
To find the scaling of $\delta E_x$ on these parameters, we fix one of them and vary 
systematically the remaining one. We find that the simulation data satisfies the scaling
\begin{equation}
 \delta E_x(x=0,y=0,s=s_0) \propto \frac{g(N_G)}{\sqrt{N_M}} \ ,
\end{equation}
where the dependence on the grid resolution, given by the function $g(N_G)$, is determined by the specific integration method of Poisson's equation. For the multi-particle tracking library \small MICROMAP\normalsize~\cite{MICROMAP}, that incorporates a spectral method, we find that $g(N_G) \simeq \sqrt[4]{N_G}$ for well resolved beams, which is used throughout the article. In particular, we find that the level of numerical noise can be kept constant, if we choose our simulation parameters such that
\begin{equation}
 \frac{\sqrt{N_G}}{N_M} = const. 
\end{equation}
A dependence of the electric field fluctuation on $x$ and $y$ is created for 
non-constant particle distributions, as the number of macro-particles per 
cell varies spatially.
For round Gaussian beams this is given by the square root of the p.d.f., as we find
\begin{eqnarray}\label{eq:scaling_law}
  \delta E_x(x,y,s) = \frac{\delta E_{x,0}\sqrt[4]{N_G}}{\sqrt{N_M}} \ e^{-\frac{x^2+y^2}{2\sigma_r^2(s_j)}} \xi(x,y,s) \ .
\end{eqnarray}
Here,
\begin{eqnarray}\label{eq:scaling_law_fac}
  \delta E_{x,0} = \delta E_x(0,0,s) \ \sqrt{\frac{N_{M,0}}{ \sqrt{N_{G,0}}}} ,
\end{eqnarray}
is a factor to quantify the standard deviation of the electric field. 
It is obtained by calculating the standard deviation in the center of the beam, $\delta E_x(0,0,s)$, for multiple random start initializations for $N_{M,0}$ macro-particles and $N_{G,0}\times N_{G,0}$ grid points.  
We normalize $\delta E_x(0,0,s)$ to $N_{G,0}$ and $N_{M,0}$, such that $\delta E_{x,0}$ is independent of these simulation parameters.

\begin{figure}[htb]
   \begin{picture}(150,190)
       \put(-110,-20){\includegraphics[scale=0.85]{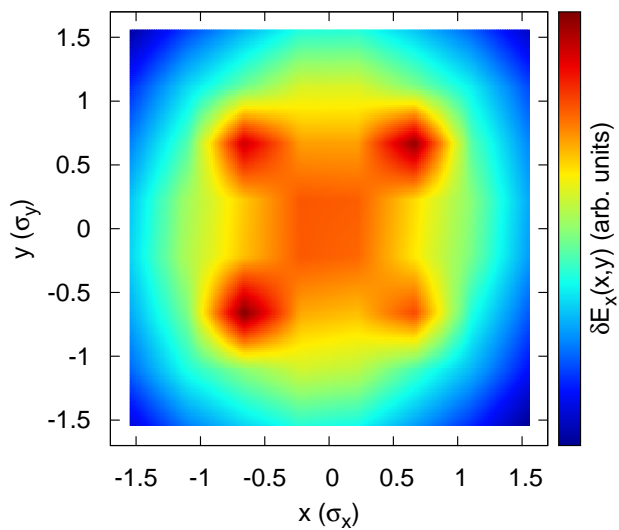}}
   \end{picture}
  \caption{(Color) Standard deviation of the transverse electric field, see Eq.~\ref{eq:scaling_law}, for a round Gaussian beam. The exponential spatial dependence of the field fluctuations is superimposed by a grid texture caused by the mesh of the PIC solver. The simulation parameters are given in the text.}
  \label{fig:standard_deviation}
\end{figure}

The factor $\xi(x,y,s)$ describes the effect due to the bi-linear interpolation of the electric field in between grid points, which causes a grid texture. As an example, the standard deviation for $N_G=16$ and $N_M=2000$ and a round Gaussian beam is given in Fig.~\ref{fig:standard_deviation}. The statistics is done with 500 random start initializations. 

The scaling law, Eq.~\ref{eq:scaling_law}, is valid as long as the beam is reasonably resolved, which is the case for $N_G\geq 16$ grid points within $\left[-2\sigma_x,2\sigma_x\right]$ (and resp. in $\left[-2\sigma_y,2\sigma_y\right]$) of the Gaussian particle distribution. The effects of a low mesh resolution were recently studied in Ref.~\cite{HofmannGrid}, where it is shown that due to grid heating the emittance growth is artificially enhanced. In the studies presented in this paper, we consider sufficiently resolved particle distributions, as this is the preferential case in space charge simulations. 

\begin{figure}[htb]
  \includegraphics[scale=0.7]{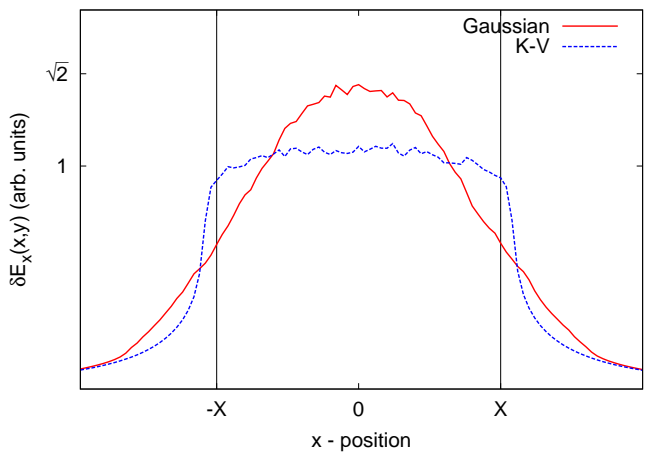}
  \caption{(Color) Electric field fluctuations $\delta E_x(x,y)$ for a  
  K~-~V distribution of size $X=2\sigma_x$ and a rms equivalent Gaussian beam. 
  Inside the beam, i.e $x\in[-X,X]$, the field fluctuations for the K~-~V 
  distributions are approximately constant.}
  \label{fig:fluctuations_KV}
\end{figure}

Using the scheme presented in this section, we can easily obtain scaling 
laws for other particle distributions. 
In the following, we study the Kapchinski-Vladimirski (K~-~V) particle 
distribution \cite{KV-distribution}, i.e. a spatially 
constant distribution of particles in the transverse plane, causing linear 
space charge forces inside the beam. For this distribution we find that 
the electric field fluctuations $\delta E_x(x,y,s=s_0)$ and $\delta E_y(x,y,s=s_0)$ inside 
the beam are constant, since the uncertainty of a PIC solver depends on the number of macro-particles per cell. 
Therefore, the electric field fluctuations inside a K~-~V beam are given by
\begin{eqnarray}\label{eq:scaling_law_KV}
  \delta E_x(x,y,s) = \frac{\delta E_{x,0}\sqrt[4]{N_G}}{\sqrt{N_M }} \ \xi(x,y,s) .
\end{eqnarray}
In Figure ~\ref{fig:fluctuations_KV}, we show the spatial dependence 
of the electric field fluctuation $\delta E_x(x,y,s=s_0)$ for a K~-~V distribution of size 
$X=2\sigma_x$ and a rms equivalent Gaussian beam \cite{rmsEq1,rmsEq2} of rms size $\sigma_x$. 
For this study we took 1000 samples of electric fields (random start), 
while the beam was well resolved with a fine grained mesh of 
$N_G\times N_G=512\times 512$ grid points. 

As seen in Fig.~\ref{fig:fluctuations_KV}, the maximum electric field fluctuations 
for the Gaussian beam are by a factor $\sqrt{2}$ larger than for a rms equivalent K~-~V beam. 
This is due to the fact, that in the center of the Gaussian beam the density is the double 
with respect to a rms equivalent K~-~V beam. The number of physical particles in the center is therefore by 
a factor $2$ larger for Gaussian beams, which enhances the electric field fluctuations.  

The results obtained in this section apply for the numerical noise 
in coasting beam simulations. However, the same method can also be used to find the dependence 
of the fluctuations $\delta E_x(x,y,s)$ for a bunched beam in the 2.5D scheme, 
see Ref.~\cite{IPAC2014}. 

\section{Noise affecting the single particle dynamics} 
\label{ch:SingleParticle}
In this section, we study the effect of a fluctuating force field on the 
dynamics of a single particle by use of the random walk theory. 
Later, these findings can be applied on a large ensemble of particles, 
such that conclusions on artificial diffusion of a full beam can 
be drawn.

\subsection{Random walk model}
A particle $i$  at the longitudinal position $s$ is described by its 
transverse phase space coordinates $x_i,x^\prime_i,y_i,y^\prime_i$. 
Its dynamics in a linear lattice is given by the Hill equations
\begin{equation}\label{eq:Hill}
  x^{\prime\prime}_i - k_x x_i=0, \qquad  y^{\prime\prime}_i - k_y y_i=0,
\end{equation}
with $k_x$, $k_y$ the restoring force terms. In the following, only 
the $x$-plane is discussed, since the $y$-plane can be treated equivalently. 

A particle $i$ with a charge $q$ and mass $m_0$ of a high intensity beam 
is exposed to a force created by the electric field $E_x(x_i,y_i,s)$ of 
the beam. The electric field is calculated in the rest frame of the beam, 
which is the standard approach in beam dynamics with space charge~\cite{ReiserBook}, 
and is used in this form throughout all discussions in this paper. 

The effect of the electric self-field on particle dynamics 
is modeled via space charge kicks \cite{GiulianoMICROMAPDescription}, 
whose effect is to change the particle coordinates $x'$:
\begin{equation} \label{eq:SCK}
  x^\prime_i (s_j) \rightarrow x^\prime_i(s_j) + 
 \Delta s \frac{q E_x(x_i,y_i,s_j)}{m_0c^2\beta^2\gamma^3},
\end{equation}
with $c$ the speed of light, and  $\beta \ $ and $\gamma$ the relativistic 
factors. 
For simplicity, the longitudinal distance between two consecutive space 
charge kicks, given by  $\Delta s = s_{j+1}-s_j$, is  kept constant during the tracking. 

By knowing position and charge state of all particles, the resulting 
space charge field on a particle located at $x_i,y_i,s_j$, given by 
$E_{x,0}(x_i,y_i,s_j)$, is calculated as a superposition of all 
Coulomb fields. We refer to this as the exact solution. 
However, if approximations are used in order to reduce the computational 
cost of the space charge field calculation, imprecisions 
are generated. See e.g. Refs.~\cite{PlasmaBook1,PlasmaBook2,PIC1,PIC2,PIC3} 
for a review. 

In the following, the effect of electric field fluctuations is studied 
in a simplified mathematical model, where the standard deviation is subtracted
or added randomly to the exact solution, i.e.
\begin{eqnarray}\label{eq:e_fluctuations}
  E_x(x_i,y_i,s_j) =  \nonumber \\ E_{x,0}(x_i,y_i,s_j) + \tilde{Z}_{ij} \delta E_x(x_i,y_i,s_j).
\end{eqnarray}
Here, $\tilde{Z}_{ij} =  1$ or $\tilde{Z}_{ij} = - 1$ is defined randomly with equal probability 
for any particle index $i$ and any longitudinal position index $j$. 
This model yields an average of $E_{x,0}(x_i,y_i,s_j)$ and a standard 
deviation of $\delta E_x(x_i,y_i,s_j)$ for a large number of electric field calculations. 

The fluctuations $\tilde{Z}_{ij} \delta E_x(x_i,y_i,s_j)$ are assumed to be statistically 
independent of each other,
i.e. for every longitudinal position $s_j$, where a space charge kick 
is applied, the fluctuation $\tilde{Z}_{ij} \delta E_x(x_i,y_i,s_j)$ does not depend 
on the fluctuation at any other longitudinal position 
$s_{j+k}=s_j+k\cdot \Delta s$ for any $k\in\mathbb{N}$. 
Throughout this article, we refer to this kind of fluctuations as 
\textit{decorrelated numerical noise}. 

Equation~\ref{eq:Hill},~\ref{eq:SCK}, and~\ref{eq:e_fluctuations} 
are the model for the evolution of a single particle affected 
by numerical noise in a high intensity beam using the standard integration 
with the one kick approximation. 
We propose that the combined effect of numerical noise and transport can be modeled 
via an effective map:
\begin{eqnarray}\label{eq:rw_mixing}
 & M\left( \begin{array}{c}x_{i}(s_j)\\ 
 x^\prime_{i}(s_j) + \Delta s \frac{q [E_{x,0}(x_{i},y_{i},s_j) + 
 \tilde{Z}_{ij}\delta E_x(x_{i},y_{i},s_j)]}{m_0c^2\beta^2\gamma^3} 
 \end{array} \right) 
 = \nonumber \\
 & =  \left( \begin{array}{c} \hat x_{i}(s_{j+1}) \\ 
 \hat x^\prime_{i}(s_{j+1}) \end{array} \right) 
 + M \left( \begin{array}{c}0 \\ 
 \tilde{Z}_{ij} \Delta x^\prime(x_{i},y_{i},s_j) \end{array} \right),
 \label{eq:rw_map}
\end{eqnarray}
with $\Delta x^\prime(x_{i},y_{i},s_j)$ the strength of the random kick, that is defined by
\begin{equation}
\small{
 \Delta x^\prime(x_{i},y_{i},s_j) = 
 \Delta s \frac{q \delta E_x(x_{i},y_{i},s_j)}{m_0c^2\beta^2\gamma^3}.
}
   \label{eq:rw_map2}
\end{equation}
The left hand side of Eq.~\ref{eq:rw_map} is the usual one step integration, 
where $M$ is the transport matrix of a particle between two consecutive 
space charge kicks obtained from Eq.~\ref{eq:Hill}. 
The coordinates $\hat x_{i}(s_{j+1}), \hat x^\prime_{i}(s_{j+1})$ are those 
of the particle transported noise-free with space charge from 
$s_j$ to $s_{j+1}$. 
The repeated application of Eq.~\ref{eq:rw_mixing} mixes the noise 
$\tilde{Z}_{ij} \Delta x^\prime(x_{i},y_{i},s_j)$ to both planes $x$ and $x^\prime$. 
Later, we model this effect by introducing an integrated effective noise kick.

To model the effect of decorrelated numerical noise on $N_M$ macro-particles 
at $N_k$ space charge kick positions, we define 
two sets of independent variables 
$Z_{ij},Z'_{ij}$ with $1\le i \le N_M$, $1\le j \le N_k$, in which 
$Z_{ij}=-1$ or $Z_{ij}=1$ (resp.  $Z^\prime_{ij}=-1$ or 
$Z^\prime_{ij}=1$) defined randomly with equal probability for any index $i$, $j$.
As suggested in Appendix \ref{ch:Appendix}, we 
describe the integrated effect of numerical noise on the phase space coordinates by using an effective model:
\begin{eqnarray}\label{eq:kick}
M \left( \begin{array}{c}0 \\ 
\tilde{Z}_{ij}^\prime \Delta x^\prime(x_{i},y_{i},s_j) 
\end{array} \right)  \nonumber
\rightarrow \\
\left( \begin{array}{c} Z_{ij}\beta_x (s_j)  \\  Z_{ij}^\prime  \end{array} 
\right) \frac{\Delta x^\prime(x_i,y_i,s_j)}{\sqrt{2}} .
\end{eqnarray}
The additional random elements account for the independence of both planes 
$x$ and $x^\prime$, while the normalization factor $1/\sqrt{2}$ stems 
directly from the mathematical model presented in Appendix \ref{ch:Appendix}. 
The effect of Equation~\ref{eq:kick} is to create a random walk 
\cite{RandomWalkBook}: At each longitudinal position $s_j$, 
where a space charge kick is applied on single particles, a 
random kick is added to both phase space coordinates.

\begin{figure}[htb]
   \begin{picture}(150,190)
       \put(-110,-20){\includegraphics[scale=0.95]{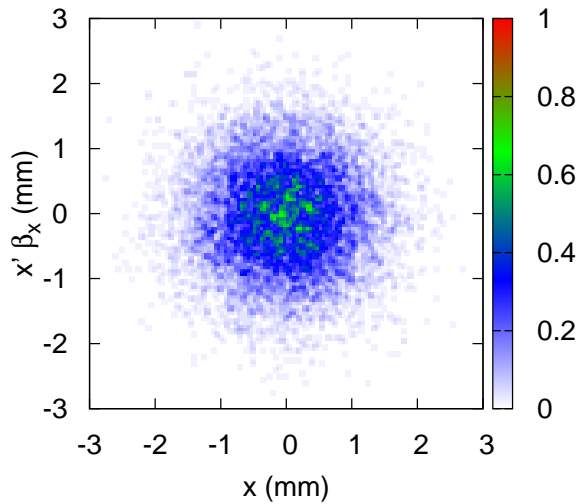}}
   \end{picture}
  \caption{(Color) Distribution of final positions of a test particle 
initialized at $x=0$, $x^\prime=0$ after tracking in a constant focusing 
channel for $n=10000$ different random initializations of macro-particles. 
 The distribution is normalized to the number of particles found in the center.}
  \label{fig:random_walk} 
\end{figure}

We test this ansatz in the following numerical experiment.
A test-particle is initialized at $x=0$ and $x^\prime=0$, while $N_M=1000$ 
macro-particles 
are distributed according to a matched Gaussian probability density function. 
The beam is tracked for 1000 turns in a constant focusing channel of length 
$L=1 \ \text{m}$ 
with a tune set to $Q_x\simeq Q_y \simeq 0.064$, while the space charge 
induced tune shift is set to $\Delta Q_x \simeq \Delta Q_y\simeq -0.005$. 
The two-dimensional Poisson solver uses a mesh of 
$N_G\times N_G = 64\times 64$ grid points. 
The integration applies one space charge kick per turn, such that 
$\Delta s=L$. The tracking is repeated $n=10000$ times for the same physical 
conditions, but with different random initializations of the particle 
distribution. 
Due to numerical noise the test particle is found at different final 
positions. 
In Figure~\ref{fig:random_walk}, we plot the distribution of final positions 
of the test particle after 1000 turns. 
The distribution is normalized to the number of particles found in the 
center of the beam, which is the maximum of the distribution.\\

The results presented in Fig.~\ref{fig:random_walk} show that numerical 
noise affects both phase space coordinates 
$x$ and $\beta_x x^\prime$ in equal measure for a sufficiently large number 
of space charge kicks. 

\subsection{Application: Single particle emittance}
In order to use the random walk model, we have to specify the spacial dependence of the fluctuations $\delta E_x(x_i,y_i,s_j)$. 
As a simplified approach we consider
\begin{equation}\label{eq:fluct_simple}
  \delta E_x(x_i,y_i,s_j)=\delta E_x(0,0) \ e^{-\frac{x_i^2+y_i^2}{2\sigma_r^2}},
\end{equation}
with  $\delta E_x(0,0,s_j)=\delta E_x(0,0)$ the electric field fluctuation 
in the center of the beam, which are, only in this first approach, assumed to be identical for any longitudinal position $s_j$. Equation ~\ref{eq:fluct_simple} can be considered as a simplified model for the electric field fluctuations of a Gaussian beam, as derived previously, see Eq.~\ref{eq:scaling_law}.

We now apply the random walk model, as given by Eq.~\ref{eq:kick}, 
on a single particle $i$, that is initialized at the phase space 
position $x_{i,0},x^\prime_{i,0}$ and tracked by using Eq.~\ref{eq:Hill} 
and~\ref{eq:SCK}. 
Following the standard theory of random walks, the most probable position 
of the particle after a series of $p$ random kicks, 
is given by the position for a noiseless tracking. Figure~\ref{fig:random_walk} confirms this, 
as the maximum of the distribution of final positions is in the center. 
After $p$ random kicks, the variances of the particle's phase space 
positions $x_{i,p},x^\prime_{i,p}$ scale as
\begin{equation}\label{eq:rw_second_order}
\begin{aligned}
  \langle x_{i,p}^2\rangle - \langle x_{i,p}\rangle^2 \propto p \delta E_x^2(0,0), \\ 
  \langle x^{\prime  2}_{i,p}\rangle - \langle x^{\prime}_{i,p}\rangle^2  \propto p \delta E_x^2 (0,0).
  \end{aligned}
\end{equation}
Each particle can be associated with a  single particle emittance
\begin{equation}\label{eq:SPE}
  \varepsilon_{SP,x_i}=\gamma_x x_i^2 +2\alpha_x x_i x_i^\prime +\beta_x x_i^{\prime 2} ,
\end{equation}
with Twiss parameters $\alpha_x, \beta_x, \gamma_x$. 
Here, we discuss the effect of decorrelated numerical noise on the initial single particle 
emittance $\varepsilon_{SP,x_i,0}$ in the context of the expectation value 
of $\varepsilon_{SP,x_i,p}$, i.e. the single particle emittance after $p$ random kicks. Assuming that numerical noise is the dominant
cause of emittance growth, we write the expectation value of $\varepsilon_{SP,x_i,p}$ as a variation of the initial single particle emittance:
\begin{equation}
 \langle\varepsilon_{SP,x_i,p}\rangle \simeq \varepsilon_{SP,x_i,0} + 
 \Delta \varepsilon_{SP,x_i,p} .
\end{equation}
Using Equations~\ref{eq:rw_second_order} and~\ref{eq:SPE}, 
we find the scaling
\begin{equation}
  \Delta \varepsilon_{SP,x_i,p} \propto p \delta E_x^2(0,0) .
\end{equation}
Thus, in average, the single particle emittance grows over many realizations of space charge kicks 
due to decorrelated random fluctuations in the space charge field. 

\section{Noise affecting a distribution of particles}\label{ch:Distribution}

In this section, we apply the random walk model on each particle of the beam. 
We then find a scaling on simulation parameters for noise induced 
diffusion. As an application, we predict the artificial emittance growth for a 
self-consistent space charge simulation for the SIS100 synchrotron 
at FAIR \cite{FAIR}.

\subsection{Effect of numerical noise on the rms-emittance}\label{Sec:scaling_law_emi}
The standard deviation of electric field fluctuations $\delta E_x(x,y,s)$ is given by  
Eq.~\ref{eq:scaling_law} and respectively Eq.~\ref{eq:scaling_law_KV} for any particle at position $x$, $y$, $s$. 
Since this result applies to any particle, it enables us to study the effect of decorrelated numerical noise 
on the full ensemble of particles, and in particular on the rms-emittance. 

In the following, we make use of the standard definition of the rms-emittance in the $x$-plane 
\begin{eqnarray} \label{eq:rms_emittance}
  \varepsilon_x^2(s) = \langle x^2\rangle \langle x^{\prime 2}\rangle -\langle xx^\prime\rangle ^2= 
  \nonumber \\ 
  = \frac{1}{N_M^2}\left[\sum_{i=1}^{N_M}x_i^2 
  \sum_{i=1}^{N_M}x^{\prime 2}_i - \left(\sum_{i=1}^{N_M} x_ix^\prime_i\right)^2\right], 
\end{eqnarray}
while the discussion for the $y$-plane is similar. 
The symbol $\langle \cdot \rangle$ denotes the average of the particle distribution. 
Let $\varepsilon_x(s_{j+1})$ be the beam emittance resulting from the application of a space charge 
kick at $s_j$ and transported by $\Delta s$ to $s_{j+1}$.  
By using the effective transport map Eq.~\ref{eq:rw_map}, and with the noise modeling of 
Eq.~\ref{eq:kick}, we predict $\varepsilon_x(s_{j+1})$ from 
the beam distribution at $s_{j}$. 
In fact, from Eq.~\ref{eq:rms_emittance} it follows
\begin{widetext}
\begin{eqnarray} \label{eq:rms_emittance_rw}
 & \varepsilon_x^2(s_{j+1}) =  
  \frac{1}{N_M^2}
  \sum_{i=1}^{N_M}
  \left(\hat x_{i}(s_{j+1}) + 
        \frac{Z_{ij}\beta_x(s_j)\Delta x^\prime(x_i(s_j),y_i(s_j),s_j)}
       {\sqrt{2}}
  \right)^2 
  \sum_{i=1}^{N_M}
  \left(
        \hat x_{i}^{\prime}(s_{j+1}) + 
        \frac{Z^\prime_{ij}\Delta x^\prime(x_i(s_j),y_i(s_j),s_j)}
                {\sqrt{2}} 
  \right)^2 
  - \nonumber \\
 & \frac{1}{N_M^2} 
  \left[ 
  \sum_{i=1}^{N_M} 
  \left(
        \hat x_{i}(s_{j+1}) + 
        \frac{Z_{ij}\beta_x(s_j)\Delta x^\prime(x_i(s_j),y_i(s_j),s_j)}
                 {\sqrt{2}}
   \right)
   \left(
         \hat x_{i}^{\prime}(s_{j+1}) + 
         \frac{Z^\prime_{ij}\Delta x^\prime(x_i(s_j),y_i(s_j),s_j)}
                 {\sqrt{2}}
   \right)
   \right]^2 ,
\end{eqnarray}
\end{widetext} 
where the coordinates $\hat x_{i}(s_{j+1}), \hat x'_{i}(s_{j+1})$ 
are those of the particle transported from $s_j$ to $s_{j+1}$ in presence 
of space charge, but without noise. In the following, we derive the artificial emittance growth averaged over many applications of space charge kicks. For this we find, that all terms proportional to $Z_{ij}$, or $Z'_{ij}$ have a minor contribution to the emittance growth, since they are in average zero with a variance proportional to the number of space charge kicks applied. The same argument holds true for the product of $Z_{ij}$ and $Z_{ij}^\prime$, since they are statistically independent. The terms proportional to $Z_{ij}^2$, and $(Z_{ij}')^2$ do not average to zero as $(Z_{ij})^2=(Z_{ij}^\prime)^2=1 \ \forall \ i,j$, and thus they dominate the contribution to the emittance growth. All terms proportional to $(\Delta x^\prime)^4$ are much smaller than any 
other term 
in Eq.~\ref{eq:rms_emittance_rw}, and can thus be neglected. 
Finally, as we consider matched beams, we use 
$\langle x^2\rangle = \langle \beta_x^2 x^{\prime 2}\rangle$, and therefore 
we find
\begin{eqnarray}\label{eq:Expanded}
   \varepsilon_x^2(s_{j+1}) \simeq \hat\varepsilon_x^2(s_{j+1}) + \nonumber \\
   \langle 
   \hat x^2(s_{j+1})\rangle \langle (\Delta x^\prime(x(s_j),y(s_j),s_j))^2 
   \rangle.
\end{eqnarray}
Here, $\hat\varepsilon_x^2(s_{j+1})$ is the square of the emittance of the beam propagated without noise.The right hand side of this equation characterizes the averaged artificial 
emittance growth in one integration step; its explicit expression is given by
\begin{widetext}
\begin{eqnarray}\label{eq:delta_E_long}
  \Delta\varepsilon_x^2(s_{j+1})=  
  \langle \hat x^2(s_{j+1})\rangle 
  \langle (\Delta x^\prime(x(s_{j}),y(s_{j}),s_j))^2 \rangle  
  = \left( \frac{q \Delta s  } {m_0c^2\beta^2\gamma^3}\right)^2  
  \langle \hat x(s_{j+1})^2\rangle 
  \langle \delta E_{x}^2(x(s_j),y(s_j),s_j)\rangle  .
\end{eqnarray}
\end{widetext}
In the following, $\delta E_x(x,y,s)$ as given by Eq.~\ref{eq:scaling_law} and 
Eq.~\ref{eq:scaling_law_KV}, is used with the simplifying condition that 
the fluctuations due to the bi-linear interpolation are set to zero, i.e. $\xi(x,y,s)=1$. 
This ansatz well describes a fine grained mesh. 

The average of $\delta E_x^2(x,y,s)$ over $x$ and $y$ can be evaluated by an integration over the particle 
distribution function, if the number of macro-particles $N_M$ is large enough to approximate the p.d.f.
Using the notation $\sigma_{r,j}=\sigma_{r}(s_j)$, we find for a Gaussian beam
\begin{eqnarray}\label{eq:fluct_G}
\langle 
 \delta E_x^2(x,y,s_j)\rangle = & 
 \int dx \int  dy f(x,y) \delta E_x^2(x,y,s_j) \nonumber \\ 
 = &\frac{\sqrt{N_G}}{N_M} 
    \frac{\delta E_{x,0}^2}{2\pi\sigma_{x,j}\sigma_{y,j}} 
    \int dx e^{\frac{-x^2}{\sigma_{x,j}^2}} \int  dy 
    e^{\frac{-y^2}{\sigma_{y,j}^2}} \nonumber \\ 
= &\frac{1}{2} \left( \frac{\delta E_{x,0}^2 \sqrt{N_G}}{N_M} \right).
\end{eqnarray}
For a K~-~V distribution we find respectively
\begin{eqnarray}\label{eq:fluct_KV}
  \langle \delta E_x^2(x,y,s_j) \rangle = 
  \frac{\delta E_{x,0}^2\sqrt{N_G}}{N_M} .
\end{eqnarray}
Using Eq.~\ref{eq:delta_E_long},~\ref{eq:fluct_G}, and~\ref{eq:fluct_KV}, we find in general 
\begin{eqnarray}
 \Delta\varepsilon_x^2(s_j)  = \Lambda \frac{\sqrt{N_G}}{N_M}
 \left( \frac{q\delta E_{x,0}}{m_0c^2\beta^2\gamma^3}\right)^2
\hat\sigma_{x,j+1}^2
 (\Delta s)^2,
 \label{eq:emittance_growth_KV_G}
\end{eqnarray}
with $\Lambda$ a coefficient that incorporates the type of distribution, with 
$\Lambda=1$ for a KV, and $\Lambda=1/2$ for a Gaussian.
We thus find a dependence of the noise induced emittance growth 
on the number of macro-particles $N_M$, the number of 
grid-points $N_G$, the integration length $\Delta s$ and the variance 
$\sigma_{x,j+1}^2$ of the distribution. 

The factor of $\Lambda=1/2$ for a Gaussian beam can be understood 
in the following way: 
For a K~-~V particle distribution, macro-particles are equally distributed, 
and the number of particles in each cell is therefore constant. 
In contrast to this, macro-particles are mostly located in the center for Gaussian beams. Therefore, 
the relative fluctuations of the electric field $\delta E_x /E_x$ are smaller, compared to a rms
equivalent K~-~V beam.

In this paper we intend to discuss the effect of PIC induced noise in a situation where the beam does not already exhibit 
emittance growth. 
Therefore, in the context of Eq.~\ref{eq:Expanded}, we find that $\hat\varepsilon_x^2(s_{j+1})$, the squared emittance of the beam propagated 
without noise, will be equal to $\varepsilon_x^2(s_{j})$.
In addition, the treatment of a constant focusing lattice further simplifies the 
formulas, as we find $\langle \hat x^2(s_{j+1})\rangle = \hat\sigma^2_{r,j+1}=\sigma^2_{r,j}$.  
Therefore, if the beam has no intrinsic emittance growth, Eq.~\ref{eq:Expanded} 
becomes 
\begin{eqnarray}\label{eq:Expanded1}
   \varepsilon_x^2(s_{j+1}) \simeq \varepsilon_x^2(s_{j}) + \nonumber \\ +
   \langle 
   \hat x^2(s_{j+1})\rangle \langle (\Delta x^\prime(x(s_j),y(s_j),s_j))^2 
   \rangle,
\end{eqnarray}
and Eq.~\ref{eq:emittance_growth_KV_G} effectively provides the 
emittance growth $\Delta\varepsilon_x/\Delta s$ due to PIC induced
decorrelated noise: 
\begin{eqnarray}
 \frac{\Delta\varepsilon_x}{\Delta s}  \simeq 
 \Lambda \frac{\sqrt{N_G}}{N_M} 
 \frac{\sigma_{x,j}^2}{2 \varepsilon_x} 
 \left( \frac{q\delta E_{x,0}}{m_0c^2\beta^2\gamma^3}\right)^2
 \Delta s.
 \label{eq:emittance_growth_KV1}
\end{eqnarray}
\subsection{Effect on the tune shift} 
The artificial emittance growth may create another disadvantage in 
terms of an artificial change of tune-shift. 
In fact, space charge forces lead to a defocussing of the beam and thus 
to a change of the betatron tune $Q_x$. 
For the K~-~V particle distribution (un-bunched) 
the associated tune shift is given by
\begin{equation}
 \Delta Q_x (t)= - \frac{r_0 N_p}{2\pi \beta^2\gamma^3}
 \frac{1}{\varepsilon_x(t)} \propto \frac{1}{\varepsilon_x(t)},
\end{equation}
where $r_0=e^2/(4\pi\epsilon_0m_0c^2)$ is the classical particle radius. 
As the initial emittance $\varepsilon_{x}(t=0)$ 
changes due to PIC induced noise, 
while the number of physical particles $N_p$ remains constant, 
the associated tune shift changes by
\begin{equation}
 \Delta Q_x(t)= \Delta Q_{x}(t=0)\frac{\varepsilon_{x}(t=0)}{\varepsilon_x(t)} .
\end{equation}
By using the expression  
$\varepsilon_x^2(t) = \varepsilon_{x}^2(t=0) + 
\Delta \varepsilon_x^2$, with $\varepsilon_{x}^2(t) >> 
\Delta\varepsilon_x^2$, we find the approximation
\begin{equation}
  \Delta Q_x(t) \simeq \Delta Q_{x}(t=0)
  \sqrt{1 -\frac{\Delta\varepsilon_{x}^2}{\varepsilon_x^2(t=0)}} ,
\end{equation}
with the PIC induced emittance growth rate $\Delta\varepsilon_x^2$ 
as derived in the previous section. 
This derivation is valid for Gaussian beams as well, and the same scaling is found. 
However, the space charge induced tune shift is the double, 
see e.g. \cite{CAS_spacecharge}.

\subsection{Application to the SIS100}
\begin{figure}[H]
 \centering
 \includegraphics[width=1.0\columnwidth]{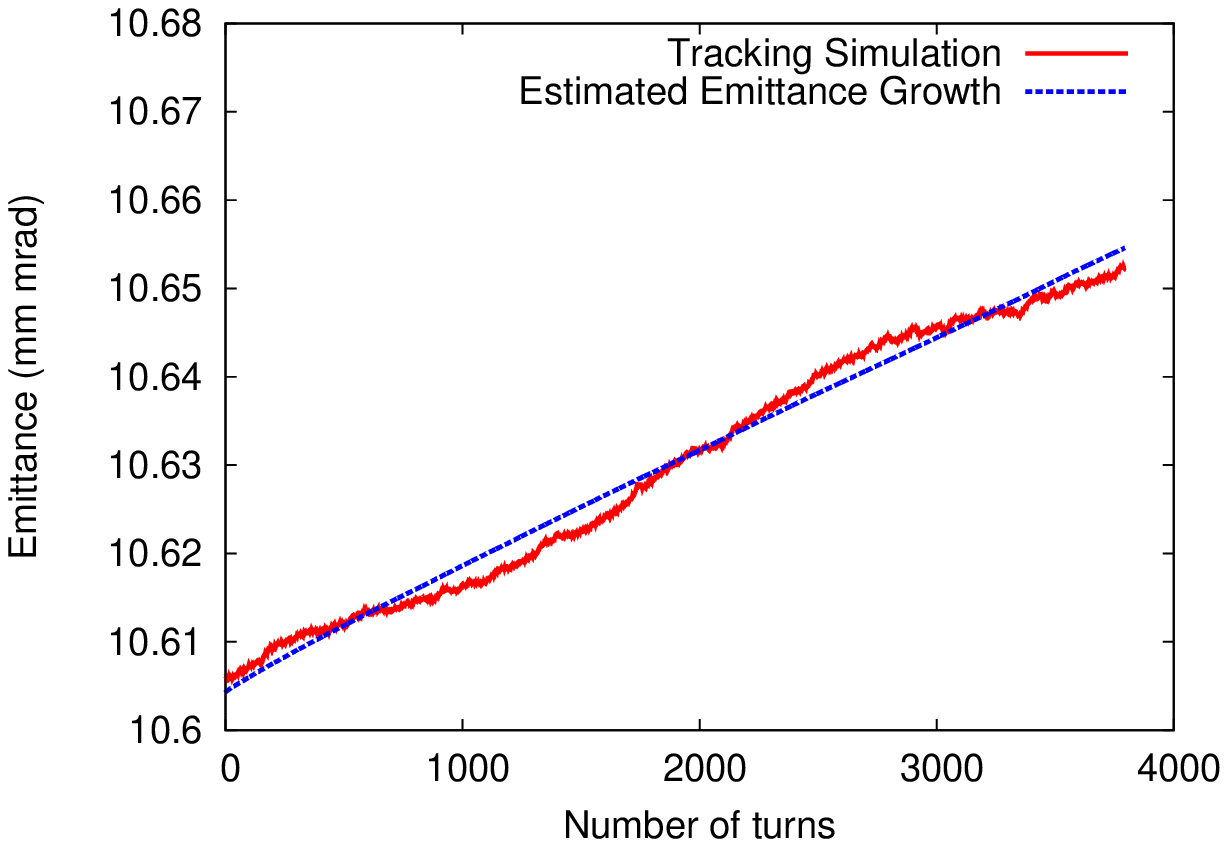} \\
 \includegraphics[width=1.0\columnwidth]{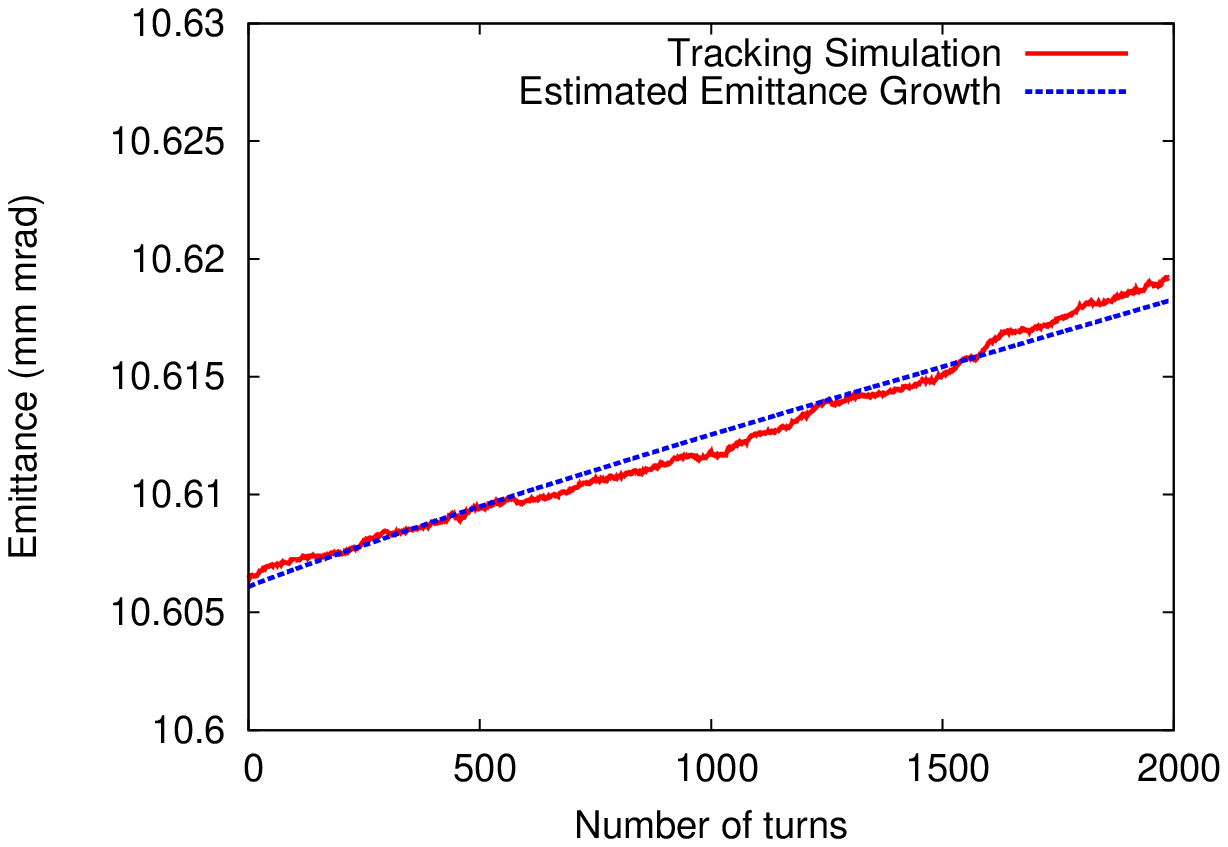}
 \caption{(Color)  
 Evolution of the emittance  
 $\varepsilon = \sqrt{\varepsilon_x^2+\varepsilon_y^2}$ due to numerical 
 noise for a simulation with $N_M=10000$ macro-particles (top) and 
 $N_M=20000$ macro-particles (bottom) for a K~-~V beam tracked in the 
 SIS100 lattice. The theory developed in this paper (blue line) predicts 
 the artificial emittance growth in the simulation (red dots). 
 The beam and simulation parameters are given in the text. }
 \label{fig:SIS100_emittance}
\end{figure}

In this section we benchmark the emittance growth predicted by 
Eq.~\ref{eq:emittance_growth_KV1} with simulations performed for the SIS100 heavy ion synchrotron~\cite{FAIR} 
(without non-linear elements). 
We use the \small MICROMAP 
\normalsize library~\cite{MICROMAP} for tracking the particles. 
The PIC solver \cite{PIC1,PIC2,PIC3} makes use of a mesh 
of $N_G\times N_G=64\times64$ grid points in a box of size $6\sigma_x\times6\sigma_y$. 
For rms-emittances of $\varepsilon_x=\varepsilon_y = 7.5 \ \text{mm} \ \text{mrad}$ the resulting space charge tune shift is $\Delta Q_x \simeq  -0.1686$ and $\Delta Q_y \simeq  -0.1693$, at tune position $Q_x\simeq 18.87 $ and $Q_y\simeq 18.72$. 
The integration step $\Delta s$ is set to 42 space charge kicks per betatron-wavelength. 
The results are presented in Fig.~\ref{fig:SIS100_emittance}, where 
the red dots show the emittance growth from the simulation, 
while the blue line is the theoretical prediction from 
Eq.~\ref{eq:emittance_growth_KV1}, i.e. the random walk model applied 
to the full beam.

For the simulations we use a coasting K~-~V beam. 
Therefore, we guarantee that space charge forces are linear and thus
PIC induced numerical noise is the dominant cause of emittance growth.

As seen in Fig.~\ref{fig:SIS100_emittance}, the theory developed 
in this section recovers the artificial emittance growth found in the 
simulation of a complex machine. 
For decorrelated numerical noise the random walk model can thus be used 
in order to 
a) understand the effect of noise on single particles, and 
b) make predictions on the evolution of the whole ensemble of particles. 

Equation~\ref{eq:emittance_growth_KV1} can be used to find reasonable 
simulation parameters $N_G$, $N_M$ and $\Delta s$. 
E.g. considering a simulation with $N_{M_1}$ macro-particles, where an artificial 
emittance growth of $\Delta\varepsilon_{x,N_{M_1}}$ is observed, we predict 
that for a simulation with $N_{M_2}$ macro-particles
\begin{equation}
  \Delta\varepsilon_{x,N_{M_2}}= 
  \frac{N_{M_1}}{N_{M_2}}\Delta\varepsilon_{x,N_{M_1}} .
\end{equation}
The scaling law, Eq.~\ref{eq:emittance_growth_KV1}, can also be used to 
find an optimal number of macro-particles for a different number of 
grid points $N_G$. In fact, choosing the number of macro-particles 
such that
\begin{equation}\label{eq:rw_equivalence}
\frac{\sqrt{N_G}}{N_M}=const,
\end{equation}
then the PIC induced emittance growth will be the same for all pairs $N_G,N_M$, 
hence one can choose the most efficient setting. 

Further, the agreement of theory and simulation allows to extrapolate the results for the SIS100 to a storage time of one second, corresponding to $2\cdot10^{5}$ turns. We find, that for the simulation setup used above, we have to use $N_M\simeq2.5\cdot10^5$ macro-particles, in order to limit artificial emittance growth to $\sim 1 \%$. If the physics case requires a higher resolution, such that we have to double the number of grid points $N_G$, we make use of Eq.~\ref{eq:rw_equivalence}. We then find, that to limit artifical emittance growth to $\sim 1\%$ for a storage time of one second, we have to use $N_M\simeq 3.5 \cdot 10^5$ macro-particles. A change of the integration length $\Delta s$ changes the slope of artificial emittance growth linearly. Thus, if we double $\Delta s$, we have to use the double amount of macro-particles to obtain the same amount of noise induced emittance growth.

For the benchmarking, we considered space charge induced tune shifts smaller than the actual design goal for the SIS100. We remark that for stronger space charge forces, the number of macro-particles has to be further increased, since the fluctuations become stronger, see Appendix~\ref{ch:Appendix_SR_Tune}. A dedicated survey for the estimation of simulation parameters for certain operational scenarios will be part of future studies.

Until now, only decorrelated numerical noise has been considered. However, the conjoint effect of the PIC algorithm and the particle tracker may cause correlation in the numerical noise. A study dedicated to this effect is presented 
in the following section.

\section{Correlated numerical noise} \label{ch:Correlations}
In the previous section we discussed the effect of decorrelated 
numerical noise on single particles and the full beam. 
In this section, we study the case where numerical noise is 
correlated. 
If a coasting beam is tracked through a circular lattice, depending 
on the integration length $\Delta s$ and the machine tunes $Q_x$ and $Q_y$, 
the particle ensemble returns (close) to its initial positions after 
a certain integer number $m^*$ of space charge kicks. 
Afterwards, the same sequence of space charge kicks is applied again, 
such as the electric field fluctuations.  
Thus, fluctuations of the electric field occur periodically 
and are no longer random. As the periodicity of the noise may create resonant effects, 
we call this a \textit{stochastic resonance} \cite{StochasticResonance}. 
These kind of correlations may lead to an enhanced emittance growth that 
exceeds the estimates of the random walk model. 

\subsection{Stochastic Resonances}
In mathematical terms, correlations in  numerical noise 
are created if 
\begin{equation}\label{eq:SR_first}
 \sum^{m^*}_{j=1} \Delta \Phi_j = 2\pi n^*, 
\end{equation} 
where $\Delta \Phi_j$ is the intra-kick phase advance in between 
$s_j$ and $s_{j+1}$ and $n^*$ is an integer. 
We define $m^*$ as the order of the stochastic resonance. 

For a constant focusing channel, where $\Delta \Phi_j = \Delta \Phi$ 
is constant for any $j$, Eq.~\ref{eq:SR_first} can be simplified to
\begin{equation}\label{eq:SR_second}
 m^*\Delta \Phi = 2 \pi n^*, 
\end{equation}
with $m^*$, and $n^*$ co-prime numbers. 
When this equation is nearly satisfied, 
the numerical noise on the first $m^*$ space charge kicks is 
completely random. 
However, for the following space charge kicks, 
the fluctuations of the electric field occur periodically 
according to 
\begin{equation}\label{eq:SR_corr}
 \delta E_x(x_i,y_i,s_j) 
 \simeq 
 \delta E_x(x_i,y_i,s_j+m^*\Delta s) ,
\end{equation} 
where, for simplicity, we assume the motion in $y$ to be frozen. 

The recurrence of fluctuations has an important property 
if $m^*$ is an even number: After $m^*/2$ space charge kicks 
the phase advance is $\pi n^*$, meaning that 
a particle $i$ with initial coordinates $x_i,y_i$ is mirrored (antisymmetry) on the phase space axes to $\overline{x}_i,y_i$. 
In this case the periodicity relation Eq.~\ref{eq:SR_corr} 
is modified, and reads
\begin{equation}\label{eq:SR_corr2}
 \delta E_x(x_i,y_i,s_j) \simeq 
  -\delta E_x\left(\overline{x}_i,y_i,
   s_j+\frac{m^*}{2}\Delta s\right).
\end{equation}
We observe that the recurrence of the electric field fluctuations 
described by Eq.~\ref{eq:SR_corr} and Eq.~\ref{eq:SR_corr2} 
is limited to a maximum number of applications $R$ of the $m^*$ space charge 
kicks. 
Afterwards correlations are canceled by accumulated noise on the particle's positions. 
The persistence of correlations $R$ depends on $m^*$ and on the granularity 
of the distribution that is controlled by $N_G$, $N_M$ and $\sigma_r$. 
Therefore, a scaling law for artificial emittance growth in the presence 
of a stochastic resonance, analogous to Eq.~\ref{eq:emittance_growth_KV_G} 
for decorrelated numerical noise, can only be derived by a dedicated study 
on the damping of correlations, which is beyond the scope of this paper.

The effect of the first $m^*$ kicks can be modeled by a random 
walk process, as described in Sec.~\ref{Sec:scaling_law_emi}. 
The order of the stochastic resonance $m^*$ fixes the number of random 
elements in $Z_{ij}$ and $Z^\prime_{ij}$. 
Thus, for a total of $N_k$  
space charge kicks in a simulation, if $m^*\geq N_k$ we retrieve 
the random walk. 
If instead $m^*$ is a small natural number, correlations in the numerical 
noise are created. Depending on the number of kicks $N_k$, the persistence of the correlations $R$, 
and the order $m^*$, correlated noise  may enhance artificial 
emittance growth. 
This issue is discussed in the context of a periodic random walk in 
Appendix \ref{ch:Appendix_RW}. 

In the following, we outline the procedure for the two types of fluctuations 
described by Eq.~\ref{eq:SR_corr}, and Eq.~\ref{eq:SR_corr2}. 
We start by considering numerical noise with a correlation as described by Eq.~\ref{eq:SR_corr}. 
We define two random variables 
$Z_{ij}, Z^\prime_{ij}$, for $1\le i \le N_M$, 
$1 \le j \le m^*$, in which $Z_{ij}=\pm 1$, and 
$Z^\prime_{ij}=\pm 1$ with equal probability. 
To model the effect of correlated numerical noise, the random elements 
$Z_{ij}$ and $Z^\prime_{ij}$ are applied periodically $R$-times on all 
particles. 

When $m^*$ is even, 
to model the noise with the property of Eq.~\ref{eq:SR_corr2},
we define two random variables 
$Z_{ij}, Z^\prime_{ij}$, for $1\le i \le N_M$, 
$1 \le j \le  m^*/2$, in which $Z_{ij}=\pm 1$, and 
$Z^\prime_{ij}=\pm 1$ with equal probability. 
In order to model the anti-symmetric property of the fluctuations, 
the next $m^*/2$ random numbers are defined by
$Z_{i, j+m^*/2} = - Z_{ij}$ and $Z'_{i, j+m^*/2} = - Z'_{ij}$, 
with $1 \le j<m^*/2$. 
This model is directly suggested by Eq.~\ref{eq:A_elements}, observing that 
after $m^*/2$ kicks the sign of $a_{ij}$, and $b_{ij}$ inverts, see Appendix~\ref{ch:Appendix}.

Therefore, $m^*$ kicks of a resonance of an even order $m^*$ have a similar
effect on a particle as the first $m^*/2$ kicks applied twice. 
This argument suggests that a resonance of order $m^*/2$ odd
will produce a similar emittance growth as a resonance of order $m^*$.

\subsection{Effect of correlations on the emittance growth}\label{ch:Correlations_EG}
\begin{figure}[htb]
 \centering
 \includegraphics[width=1.0\columnwidth, angle=0]{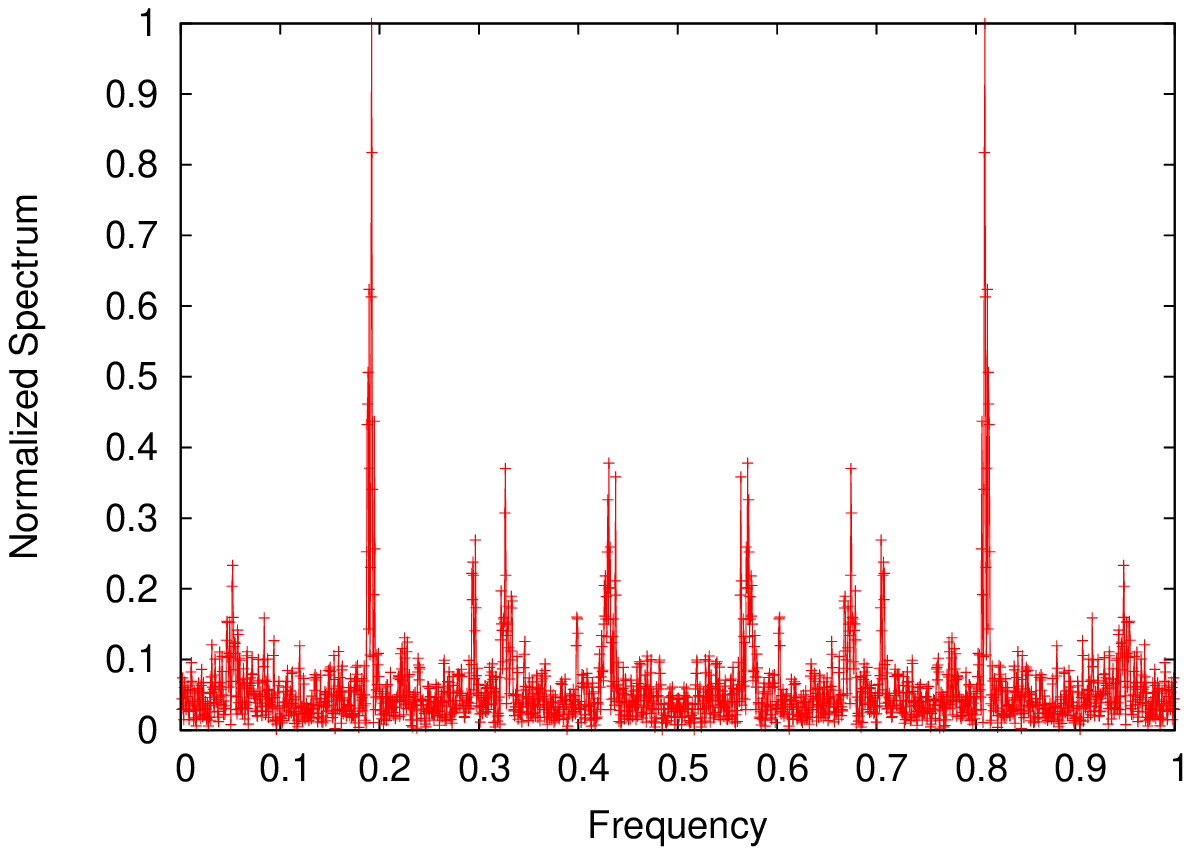}
 \includegraphics[width=1.0\columnwidth, angle=0]{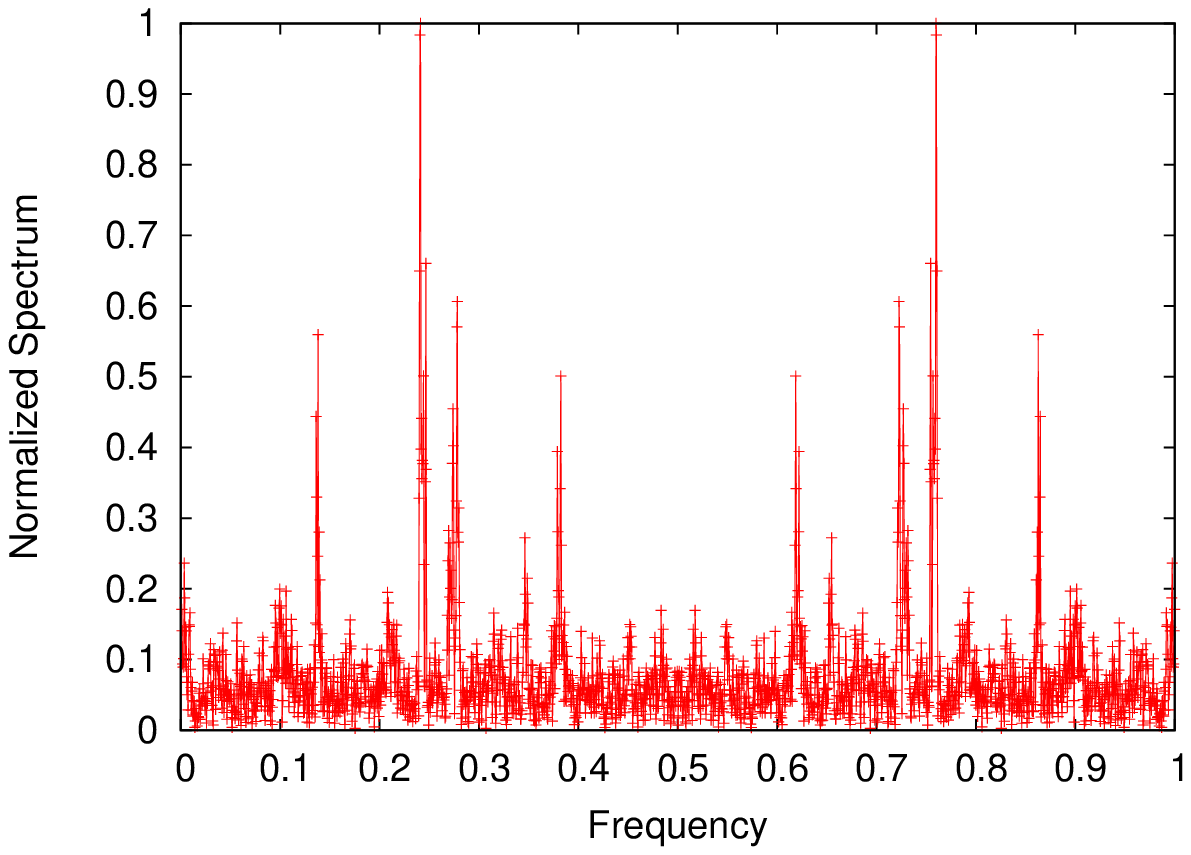}
 \caption{(Color) Spectrum analysis of the electric field in $x$-direction (top) and $y$-direction (bottom) at $x=y=0$. For this analysis, the beam is tracked in a constant focusing channel with tune $Q_x=0.19$ and $Q_y=0.24$. Peaks occur at the tune frequencies, while a coupling of both planes excites many other smaller local peaks.}
 \label{fig:spectrum_example}
\end{figure}

In the following, we study the properties of the correlations in electric field fluctuations.
For this purpose, a coasting beam is tracked in a constant focusing channel 
with machine tunes $Q_x\simeq 0.19$ and $Q_y\simeq 0.24$. 
The length of the channel is $L=3$ m. 
For simplicity, we chose the integration length $\Delta s=L$. 
The electric field $E_x(x,y,s)$ and $E_y(x,y,s)$ 
at $x=y=0$ is calculated each turn for a total of 1024 turns, 
and a spectrum is retrieved by performing a Fourier transform of the data. 
In the corresponding spectrum, see Fig.~\ref{fig:spectrum_example}, the 
highest peak occurs at the machine tune position. Other frequencies are 
excited due to a coupling of both planes. 

\begin{figure}[htb]
 \centering
 \includegraphics[width=1.0\columnwidth, angle=0]{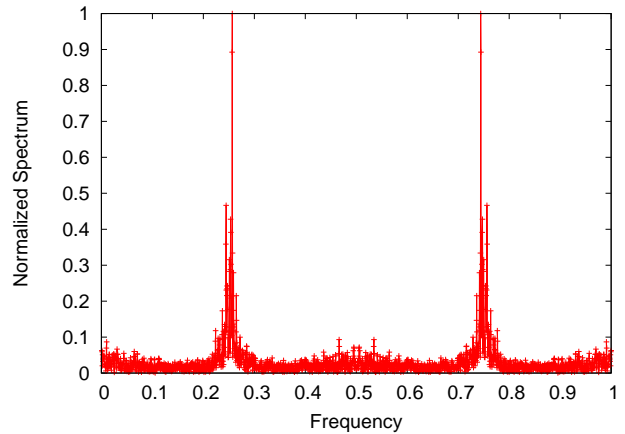}
 \caption{Spectrum analysis of the electric field in $y$-direction at $x=y=0$. The vertical tune is $Q_y=0.267$ while the horizontal tune $Q_x\simeq 1.001$ is close to an integer value. A peak occurs at the tune position, while other frequencies are no longer excited.}
 \label{fig:spectrum_clean}
\end{figure}

Since we pursue to study the effect of stochastic resonances in one plane, 
we have to ensure that we do not have any interference of the other plane. 
This can be guaranteed by setting one of the tunes close to an integer value.
If we choose $Q_x\simeq 1.001$, we decouple sufficiently the $x$ and $y$ planes, see Fig.~\ref{fig:spectrum_clean}.

\begin{figure}[htb]
 \includegraphics[width=1.0\columnwidth, angle=0]{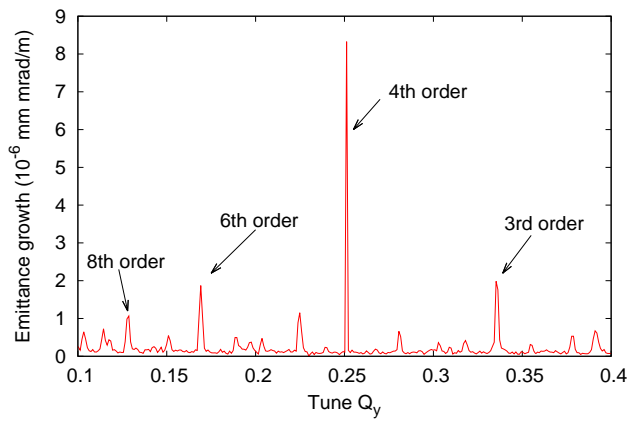}
\caption{(Color) Emittance growth rate for different vertical tunes $Q_y$, while $Q_x\simeq 1.001$. The tune $Q_y$ is changed systematically in steps of $\Delta Q=0.001$.}
 \label{fig:final_scan}
\end{figure}

In order to study the effect of the intra-kick phase advance on the artificial emittance growth,
we use again the constant focusing channel with the same simulation parameters as above.
We now systematically scan the vertical tune $Q_y$ in the range $Q_y=0.1...0.4$, 
while keeping $Q_x\simeq 1.001$ fixed. The initial emittance in the 
$y$-plane is chosen for any machine tune such that we have a round beam 
($\sigma_x=\sigma_y$). For each machine tune a coasting beam is tracked 
for 60000 turns and the slope of emittance growth per meter is calculated. 
The dependence of emittance growth on the machine tune, and thus on the 
excitation of stochastic resonances, is retrieved and shown 
in Fig.~\ref{fig:final_scan}. 

By studying the simplified scenario of a constant focusing channel, 
we guarantee a constant phase advance in between space charge kicks. 
Then, stochastic resonances are excited when a multiple of the intra-kick 
phase advance is an integer of $2\pi$, see Eq.~\ref{eq:SR_second}. 
For a realistic model of a machine the scenario gets more complex, 
as described by Eq.~\ref{eq:SR_first}. 
The emittance growth rates, as presented in Fig.~\ref{fig:final_scan}, confirm the prediction given by the theory developed in this paper. In the absence of stochastic resonances (decorrelated numerical noise), we find a growth rate that can be predicted by the random walk model. The emittance growth rate has a (local) peak whenever a stochastic resonance is hit. The lower the order of the resonance $m^*$, the stronger we see its effect on the emittance growth rate.
Further, we find that if $m^*/2$ is an odd integer, the growth rate is similar to the one of $m^*$, e.g. in the case $m^*=6$.

Until now, the space charge tuneshift was fixed for each simulation. 
In Appendix \ref{ch:Appendix_SR_Tune} we find that the width of the stopband of a stochastic resonance and the
slope of artificial emittance growth are increased for more intense beams. 

\subsection{Change of integration length}
For long-term tracking studies it is desirable to avoid the occurrence 
of correlated noise to a) minimize artificial emittance growth and 
b) avoid uncontrollable artifacts, like tails in the distribution. 
In the following, we present a strategy to ensure that the numerical noise 
is decorrelated and thus controllable by a proper choice of simulation 
parameters, as discussed above. 

\begin{figure}[htb]
  \includegraphics[width=1.0\columnwidth, angle=0]{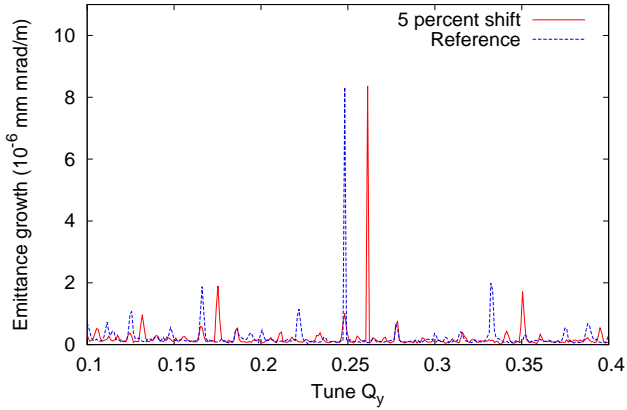}
  \caption{(Color) Emittance growth rate for different vertical tunes $Q_y$, while $Q_x\simeq 1.001$. The tune $Q_y$ is changed systematically in steps of $\Delta Q=0.001$.. The integration length is changed, compared to the blue reference line, by a factor of 0.95 (red line).}
  \label{fig:shift_scan}
\end{figure}

In most simulation studies the machine tune $Q_y$ is a multiple of the 
intra-kick phase advance, i.e.
\begin{equation}
 \sum_{i=1}^m \Delta \Phi_i = 2\pi Q_y,
\end{equation}
because the integration length $\Delta s$ is a fraction of the machine 
length $L$. For simulations close to a machine resonance, i.e. of 
type $n Q_y = N \in \mathbb{N}$ for $n\in\mathbb{N}$,  stochastic 
resonances due to PIC noise will then be excited simultaneously and will afflict the simulations. 
However, this situation can be avoided by choosing an appropriate integration 
length $\Delta s$. 
As a demonstration of the influence of the integration length on the 
occurrence of stochastic resonances, the study presented in 
Sec.~\ref{ch:Correlations_EG} is repeated for an integration length 
$\Delta s = L \cdot 0.95$. The results are presented in 
Fig.~\ref{fig:shift_scan}. 

The blue line in Fig.~\ref{fig:shift_scan} is the same as in Fig.~\ref{fig:final_scan}, where $\Delta s= L$, while the red line corresponds to $\Delta s= L\cdot0.95$. The peaks due to stochastic resonances are shifted for the 
modified integration length. 

The effect of changing the integration length can most easily be understood for the simplified case of a constant focusing channel. 
In fact, if the integration length is set to a fraction $f$ of the initial one, i.e. $\Delta s= L \cdot f$, the intra-kick phase advance is changed by a factor $f$. Therefore, the resonance condition, Eq.~\ref{eq:SR_second}, is fulfilled for another tune $Q_y^\prime$ given by
\begin{equation}
  m^*Q_y^\prime= \frac{n^*}{f} \ .
\end{equation}
We can thus effectively avoid correlated noise in our working point 
$Q_y$ by changing the integration length, and hence
guarantee that single particles are solely affected by white 
noise. Then, artificial noise effects can be mitigated for an optimal parameter 
setting obtained using Eq.~\ref{eq:emittance_growth_KV1}.

\section{Conclusion}\label{ch:Conclusion}
In this paper we developed a model for predicting the effect of white noise on a particle beam by using an equivalent effective field fluctuation acting on all phase space coordinates. We use simulation results to infer the mathematical modeling of electric field fluctuations on a single particle in the beam, and use this model to develop a treatment of the full beam ensemble. The inferred scaling law as a function of the simulation parameters enables us to estimate artificial emittance growth. 

We find that for specific machine tunes, which recall those of a machine resonance, PIC noise becomes correlated. The effect of these correlations on emittance growth is more dramatic, and we explain this unusual dynamics in terms of stochastic resonances. Our studies for the constant focusing channel show that the intra-kick phase advance plays a crucial role on numerical noise as it defines the tunes at which stochastic resonances appear.  
This understanding is used to develop a strategy to avoid or minimize correlations in numerical noise, hence avoiding stochastic resonances. The theory developed in this paper can thus be applied to control unwanted noise effects in tracking simulations of high intensity beams. 

Last, we remark that in our studies we decoupled the PIC induced numerical 
noise in the transverse planes by setting the horizontal tune $Q_x$ close to 
an integer value. However, in more realistic applications, the effect of 
coupling cannot be neglected, and certainly will introduce corrections. 
Investigations on the coupling between the planes will be part of future 
studies.

\section{Acknowledgments}
The authors acknowledge valuable remarks of Prof. Dr. I. Hofmann (GSI). The research leading to these results has received funding from the European Commission under the FP7 Research Infrastructures Project EuCARD-2, Grant Agreement No. 312453

\appendix
\section{Effect of the intra-kick phase advance on numerical noise}\label{ch:Appendix}
In the following, the combined effect of linear tracking and numerical noise is studied. For simplicity, we discuss the constant focusing channel. 

Let us consider a single particle initialized at position 
$x_i, x_i^\prime$, and track it including the space charge 
through a constant focusing channel. 
Due to numerical noise, a fluctuation  
$\Delta x^\prime(x_i,x_i^\prime,s_j)$ is induced at 
any longitudinal position $s_j$, where a space charge kick is applied. 
In total $N_k$ space charge kicks are applied, 
i.e. the particles are tracked for a distance $d= \Delta s N_k$. 
The numerical noise at position $s_j$ affects the particle's phase space 
coordinates at $s_{N_k}$ by
\begin{equation}\label{eq:A_basic}
 M^{N_k-j}  \left(\begin{array}{c} 0\\ 
\tilde{Z}^\prime_{ij}\Delta x^\prime(x_i,x_i^\prime,s_j)  \end{array} \right).
\end{equation}
We call the first component of this vector $a_{ij}$ and the second $b_{ij}$, which can be interpreted as random kicks on $x_i$ resp. $x_i^\prime$ due to numerical noise at $s=s_j$. Since we specialize on a constant focusing channel, we can decompose the transfer matrix $M$ into $M=T^{-1}R(\Delta \Phi)T$ with
\begin{equation}
 T = \left(\begin{array}{cc} \frac{1}{\sqrt{\beta_x}} & 0 \\ 0 & \sqrt{\beta_x} \end{array} \right),
\end{equation}
and
\begin{equation}
 R(\Delta \Phi)=  \left(\begin{array}{cc} \cos(\Delta \Phi) & \sin(\Delta \Phi) \\ -\sin(\Delta \Phi) & \cos(\Delta \Phi)\end{array} \right).
\end{equation}
Using $l=N_K-j$, we simplify Eq.~\ref{eq:A_basic} to
\begin{equation}
  R\left(\Delta \Phi l\right)T \left(\begin{array}{c} 0\\ 
 \tilde{Z}^\prime_{ij}\Delta x^\prime(x_i,x_i^\prime,s_j)  \end{array} \right) = T\left( \begin{array}{c} a_{ij} \\ b_{ij}  \end{array}  \right).
\end{equation}
The random elements $a_{ij}$, $b_{ij}$ can be described by
\begin{equation}\label{eq:A_elements}
\begin{aligned}
 a_{ij}=\tilde{Z}^\prime_{ij}\sin\left(\Delta \Phi l\right)\beta_x\Delta x^\prime(x_i,x_i^\prime,s_j) , \\
  b_{ij}=\tilde{Z}^\prime_{ij}\cos\left(\Delta \Phi l\right)\Delta x^\prime(x_i,x_i^\prime,s_j).
  \end{aligned}
\end{equation}
A test-particle initialized at $x_i(s_0),x_i^\prime(s_0)$ and tracked through the constant focusing channel accumulates the random elements  $a_{ij}$, $b_{ij}$. For $N_k$ space charge kicks the particle coordinates are 
\begin{equation}\label{eq:A_ab}
\begin{aligned}
 x_i(s_{N_k}) = \hat{x}_i(s_{N_k}) + \sum_{j=1}^{N_k} a_{ij} 
          = \hat{x}_i(s_{N_k}) +  a_i , \\
 x_i^\prime(s_{N_k}) = \hat{x}_i^\prime(s_{N_k}) + \sum_{j=1}^{N_k} b_{ij} 
                = \hat{x}_i^\prime(s_{N_k}) + b_i, 
 \end{aligned}
\end{equation}
where $\hat{x}_i(s_{N_k}),\hat{x}_i^\prime(s_{N_k})$ are the final coordinates 
of the test particle for a noise-free tracking. The random elements $a_{ij}$, $b_{ij}$, as derived in Eq.~\ref{eq:A_elements}, depend on the intra-kick phase advance $\Delta \Phi$. 
For most intra-kick phase advances $\Delta \Phi$ the cumulated numerical noise remains small, since the contributions $a_{ij}$ resp. $b_{ij}$ are averaged to zero, i.e.  $\langle a_i \rangle \approx 0$ and resp. $\langle b_i \rangle \approx 0$. 
We note that $a_i$ (resp. $b_i$) is a new random variable, which is the result of 
the summation of the $N_k$ random variables $a_{ij}$ (resp.  $b_{ij}$). 
The random variable $a_i$ has a variance of 
\begin{equation}
  \sigma^2_{a_{i}} = \sum_{j=1}^{N_k} \sigma^2_{a_{ij}},
\end{equation}
because the random variables $a_{ij}$ are decorrelated. 
A direct summation of this equation yields 
\begin{equation}
  \sigma^2_{a_{i}} = N_k 
  \left[
    \frac{1}{\sqrt{2}}\beta_x
    \Delta x^\prime
  \right]^2 ,
\end{equation}
which retrieves the property of the random walk. 
Therefore, we can substitute each $a_{ij}$ with the random kick  
\begin{equation}
  a_{ij} \rightarrow 
  Z_{ij}
  \frac{1}{\sqrt{2}}\beta_x \Delta x^\prime,
\end{equation}
with $Z_{ij} =\pm 1$ random, 
to retrieve the same final statistical properties after $N_k$ kicks. 
The same argument applied to the quantity $b_i$ shows that 
each $b_{ij}$ can be substituted with 
\begin{equation}
  b_{ij} \rightarrow 
  Z'_{ij}
  \frac{1}{\sqrt{2}}
  \Delta x^\prime,
\end{equation}
with $Z^\prime_{ij}=\pm 1$ random and statistically independent from the set $Z_{ij}$. 
This explains the effective random walk model used in Eq.~\ref{eq:kick}. 

\begin{figure}[htb]
 \includegraphics[width=1.0\columnwidth, angle=0]{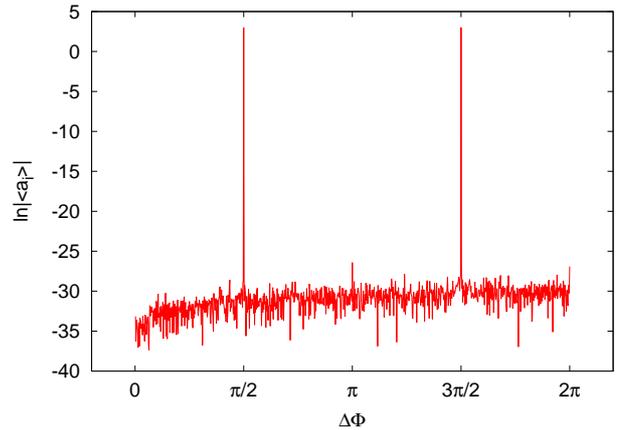}
\caption{(Color) Averaged cumulated numerical noise $\langle a_i \rangle$ for 100 initializations of $\tilde{Z}^\prime_{ij}$ following Eq.~\ref{eq:A_cor}. Peaks appear at phase advances $\Delta \Phi=\pi/2$, and $\Delta \Phi=3\pi/2$.}
 \label{fig:sum_a}
\end{figure}

\begin{figure}[htb]
 \includegraphics[width=1.0\columnwidth, angle=0]{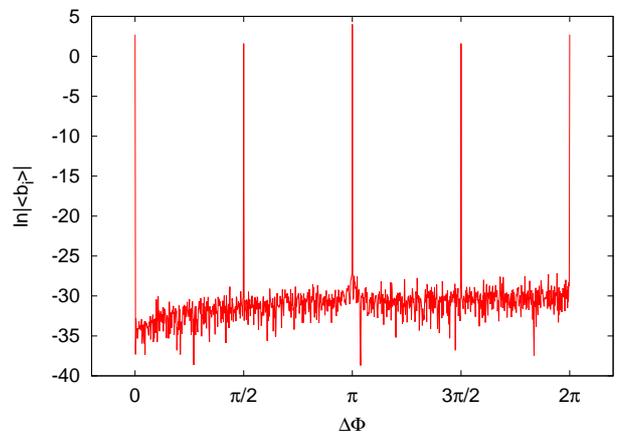}
\caption{(Color) Averaged cumulated numerical noise $\langle b_i \rangle$ for 100 initializations of $\tilde{Z}^\prime_{ij}$ following Eq.~\ref{eq:A_cor}.  Peaks appear at phase advances $\Delta \Phi=0.0$, $\Delta \Phi=\pi/2$, $\Delta \Phi=\pi$, $\Delta \Phi=3\pi/2$, and $\Delta \Phi=2\pi$.}
 \label{fig:sum_b}
\end{figure}

The random walk approach is not applicable anymore, if
\begin{equation}\label{eq:A_SR}
  m^*\Delta\Phi= 2 \pi n^*, 
\end{equation}
because a correlation is created. 
We show the breaking down of the random walk approach with a numerical experiment. For this, we define a sequence of random numbers $\tilde{Z}_{ij}$ with the correlation 
\begin{eqnarray}\label{eq:A_cor}
 \tilde{Z}_{ij}=\tilde{Z}_ {i(j+4)} \hspace{0.5cm} 1 \le j \le N_k-4,
\end{eqnarray}
which resembles correlated numerical noise in the presence of a stochastic resonance. 
The cumulated effect for a constant fluctuation $\Delta x^\prime$ for $N_k=1000$ is given by $a_i$ and $b_i$, as described in Eq.~\ref{eq:A_ab}. 
The averaged cumulated numerical noise $\langle a_i \rangle$ and 
$\langle b_i \rangle$ for 100 random configurations of $Z^\prime_{ij}$ 
following Eq.~\ref{eq:A_cor}, is shown in Fig.~\ref{fig:sum_a} and 
Fig.~\ref{fig:sum_b} for intra-kick phase advances $0 < \Delta \Phi < 2\pi$. 
\\
\\
We observe the cumulation of numerical noise for certain intra-kick phase advances, i.e. $|\langle a_i \rangle| >> 0$, 
and $|\langle b_i \rangle|>> 0 $. 
These results can be interpreted in the following way. If the condition for a stochastic resonance, see Eq.~\ref{eq:A_SR}, is not fulfilled, then the elements $a_{ij}$, $b_{ij}$ are random and thus the random walk model can be applied. Since both parameters are decoupled and are in average equivalent, the model given by Eq.~\ref{eq:kick} is justified. However, in the presence of correlations this model is not valid anymore and the noise has to be treated differently, see Sec.~\ref{ch:Correlations}.

\section{Periodic random walk}\label{ch:Appendix_RW}
To explain the effect of a periodic random walk, we study the following 
simplified case. A single particle with the one dimensional coordinate $x$ 
is affected by $N_k$ random kicks of strength $\Delta x = 1$ in arbitrary 
units. We further enforce a periodicity of $m^*$, i.e. we define a set of 
random numbers $X_i\in \{-1,1\}$ with $i\in \{1,2, \dots  m^*\}$ that is 
applied $R$ times. 
The number of repetitions $R$ is chosen to fulfill $N_K=m^*R$, with $N_K$ constant. Since the averaged position after many random kicks is $\langle x \rangle \approx 0$, the averaged second order moment can be associated with the variance. 
 
\begin{figure}[htb]
 \includegraphics[width=1.0\columnwidth, angle=0]{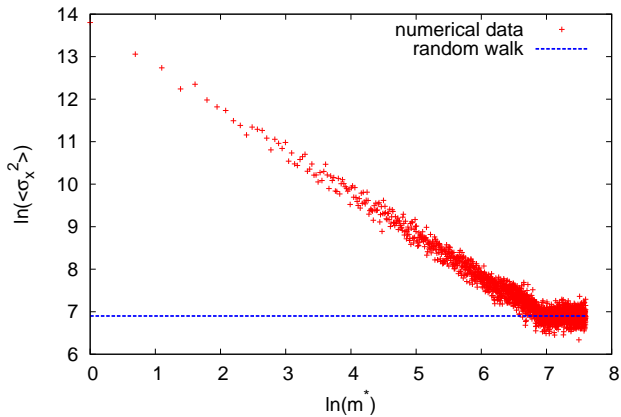}
\caption{(Color) Averaged second order moment $\langle \sigma_x^2\rangle$ after $N_k=1000$ random kicks for different periodicities $m^*$. The blue line indicates the second order moment for the non-periodic random walk (without repetition).}
 \label{fig:periodic_walk}
\end{figure}

The dependence of $\langle \sigma_x^2\rangle$ on $m^*$ is retrieved in a numerical experiment. For this purpose, $\sigma_x^2$ is evaluated for 100 random initializations of $X_i$ for each $m^* \in \{1,2,\dots 2000\}$. The results are presented in Fig.~\ref{fig:periodic_walk}.

Fitting the numerical data, we find for $m^*<N_k$ the scaling of the averaged second order moment on $m^*$
\begin{equation}\label{eq:periodic_walk}
 \langle \sigma_x^2 \rangle \propto \frac{1}{m^*}.
\end{equation}
For $m^*\geq N_k$, the averaged second order moment becomes independent of $m^*$ since we apply only $N_k$ kicks in this example. These results show that for a periodic random walk the diffusion is enhanced. The smaller the periodicity $m^*$, the larger is the diffusion. 
 
When applying these results to the situation of a stochastic resonance created by PIC tracking, we have to consider that correlations in the numerical noise are washed out due to the numerical noise itself. The scaling, as given in Eq.~\ref{eq:periodic_walk}, may suggest that the periodic random walk could be an upper bound on the diffusion, because no loss of memory is considered. If the correlations are completely washed out, the diffusion type will the one of a random walk. This defines the lower bound on the diffusion, as indicated by the blue line in Fig~\ref{fig:periodic_walk}.\\

\section{Effect of the strength of space charge on the excitation of stochastic resonances}\label{ch:Appendix_SR_Tune}
\begin{figure}[htb]
 \includegraphics[width=1.0\columnwidth, angle=0]{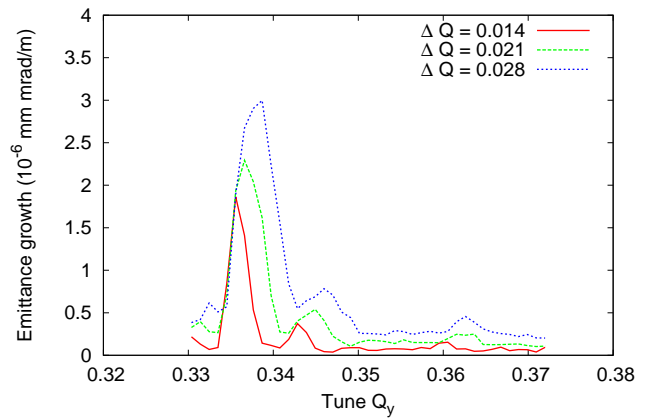}
\caption{(Color) Emittance growth rate for different vertical tunes $Q_y$ in the vicinity of the third order stochastic resonance, while the horizontal tune is $Q_x\simeq 1.001$. The vertical tune $Q_y$ is changed systematically in steps of $\Delta Q=0.001$. The colors represent different beam currents.}
 \label{fig:scan_detail}
\end{figure}

The excitation of stochastic resonances is certainly affected by the strength of the space charge force. To gain a quantitative knowledge, we repeat the study presented in Sec.~\ref{ch:Correlations_EG} for different beam currents and thus different space charge forces. The artificial emittance growth is investigated in the vicinity of the third order stochastic resonance, which is presented in Fig.~\ref{fig:scan_detail}. 

As we learn from Fig.~\ref{fig:scan_detail}, an increase of the beam , and thus the space charge force, enhances artificial emittance growth and leads to a broadening of the stop-band of the stochastic resonance. The increase of artificial emittance growth can be explained by the increase of the driving force of the resonance, i.e. of the space charge. At the same time, by increasing the space charge force we enlarge the tune spread of the beam. Therefore, the range of tunes $Q_y$  affected by the stochastic resonance grows larger.

\end{document}